\documentclass[pra,twocolumn,showpacs,preprintnumbers,amsmath,amssymb]{revtex4}
\usepackage{graphicx}% Include figure files
\usepackage{dcolumn}% Align table columns on decimal point
\usepackage{bm}% bold math

\begin{document}
\title{Finite Superposition Solutions for Surface States in a Type of Photonic Superlattices}

\author{Qiongtao Xie$^{1,2}$}
 \altaffiliation{Electronic address: xieqiongtao@yahoo.cn}

\author{Chaohong Lee$^{2,3}$}
 \altaffiliation{Electronic address: chleecn@gmail.com}

\affiliation{$^{1}$Department of Physics and Key Laboratory of Low-Dimensional Quantum Structure and Quantum Control of Ministry of Education, Hunan Normal University, Changsha 410081, China}

\affiliation{$^{2}$State Key Laboratory of Optoelectronic Materials and Technologies, School of Physics and Engineering, Sun Yat-Sen University, Guangzhou 510275, China}

\affiliation{$^{3}$Nonlinear Physics Centre, Research School of Physics and Engineering, Australian National University, Canberra ACT 0200, Australia}

\date{\today}

\begin{abstract}

We develop an efficient method to derive a class of surface states in photonic superlattices. In a kind of infinite bichromatic superlattices satisfying some specific conditions, we obtain  a finite portion of their in-gap states, which are superpositions of finite numbers of their unstable Bloch waves. By using these unstable in-gap states, we construct exactly several stable surface states near various interfaces in photonic superlattices. We analytically   explore the parametric dependence of these exact surface states. Our analysis provides an exact demonstration for the existence of surface states and would be also helpful to understand surface states in other lattice systems.

\end{abstract}


\maketitle

\section{Introduction}

Surface states are a kind of localized states found at the interfaces between two different media. In 1932, Tamm predicted that electronic surface waves may exist in a semi-infinite
one-dimensional repulsive  Kronig-Penney (KP) model~\cite{Tamm} and these surface states were named as Tamm states later. Similar results also exist in the attractive KP model~\cite{Grimley}. The energies of Tamm states lie in the forbidden gaps for the corresponding infinite KP model. The surface states depend crucially on surface termination of periodic potentials~\cite{Shockley, Stats,Lippmann}. It has been demonstrated that Tamm states appear when the periodic potentials are asymmetrically terminated and Shockley states appear while the periodic potentials are symmetrically terminated~\cite{Shockley}. Surface states have been observed in several experimental systems such as semiconductor superlattices~\cite{Ohno, Miller, Zahler,Bellessa} and magneto-photonic structures~\cite{Goto}.

In the past few decades, surface states have been studied extensively due to their potential applications in optoelectronic devices. However, an analytical demonstration of the existence of exact surface states is still absent~\cite{Heinea, Forstmann,Levinea}. For an example, to construct a surface state in the semi-infinite one-dimensional KP model, one need to calculate analytically or numerically a Bloch wave of a complex wave number and then match it with an exponentially decaying state inside the surface potential~\cite{Heinea, Forstmann, Levinea}. In a semi-infinite sinusoidal potential~\cite{Levinea} or a semi-infinite KP potential~\cite{Tamm, Grimley, Steslicka}, the energies and the wave numbers for the surface states are determined implicitly by solving a transcendental equation. In addition, there are other approximation methods for constructing surface states~\cite{Levineb}, such as the coefficient method, the scattering method, the determinant method and the integral method. Up to now, it is still a great challenge to find exact solutions for surface states~\cite{Yu, Sy, Malkovab}.

Due to the temporal evolution of quantum systems can be mapped into the spatial propagation of light waves, the engineered photonic lattices provide a highly controllable platform for exploring similar surface states in periodic quantum systems. Surface states near an interface between a periodic layered medium and a homogeneous medium are found to be analogous to electronic surface states in crystals~\cite{Yeh}. The formation of Shockley-like surface states in an optically induced semi-infinite photonic superlattice has been experimentally demonstrated~\cite{Malkovaa}. The formation of Tamm states at the boundary between two periodical dielectric structures has been reported~\cite{Kavokin}. It has been demonstrated that the nonreciprocality of the surface modes can be induced by the violation of periodicity and the violation of the time reversal symmetry~\cite{Khanikaev}. The recent advances on surface states in photonic crystals are reviewed in~\cite{Lisyanskii}.

In this paper, for a kind of infinite bichromatic superlattices satisfying some certain conditions, we  find a set of the in-gap states in superpositions of finite numbers of unstable Bloch waves. These unstable in-gap states are then used to  construct stable surface states for several typical systems of surface states, such as the semi-infinite photonic superlattice, the finite photonic superlattice, two directly coupled photonic superlattices, and two indirectly coupled photonic superlattices. The conditions for the existence of these  surface states are derived. We find that the symmetry and the existence of the surface states depend crucially on both lattice parameters and interface parameters. Our analytical results of surface states provide an optional benchmark for understanding surface waves in lattice systems.

\section{In-gap states in photonic superlattices} 

We consider the light propagation in a one-dimensional
waveguide array. Assuming the waveguide array is aligned along the
$X$ direction and the light is localized along the $Y$ direction,
the propagation of the light electric field $E(X,Z)$ along the $Z$
direction is described by an effective two-dimensional wave
equation~\cite{Della},
\begin{equation}
i\frac{\lambda}{2\pi}\frac{\partial E}{\partial Z}=-\frac{\lambda^2}
{8\pi^2 n_{s}}\frac{\partial^{2} E}{\partial X^2}+U(X)E, \label{eq1}
\end{equation}
where $\lambda$ is the free-space light wavelength and $n_s$ is the
substrate refractive index. The profile of the effective refractive
index is in form of $U(X)=[n_s^2-n^2(X)]/(2n_s)\simeq n_s-n(X)$ with
the refractive index $n(X)$ for the waveguide array. By using the
periodic modulation techniques of $n(X)$~\cite{Shandarova}, one can
build a bichromatic superlattice of $U(X)=n_1[1-\cos(\frac{2\pi
X}{\Lambda})]+ n_2[1+\cos (\frac{\pi X}{\Lambda}+\theta)]$ with the
amplitudes $n_1$ and $n_2$, the modulation period $2\Lambda$ and the
relative phase $\theta$.

By introducing two scaled variables $x=\pi X/\Lambda$ and $z=\pi \lambda Z/(4\Lambda^2n_s)$ and a transformation $\phi(x,z)=E(x,z)\exp[i(n_1'+n_2')z]$ with $n_1'=8\Lambda^2n_s n_1/\lambda^2$ and $n_2'=8\Lambda^2n_s n_2/\lambda^2$, the system is then described by
\begin{equation}
i\frac{\partial \phi}{\partial z}=-\frac{\partial^{2}
\phi}{\partial x^2}+V(x)\phi, \label{eq3}
\end{equation}
with $V(x)=-n_1'\cos(2x)+n_2'\cos(x+\theta)$. Considering its
stationary states, $\phi(x,z)=\psi(x)\exp[-i\beta z]$, the amplitude
$\psi(x)$ obeys a time-independent equation
\begin{equation}
\beta\psi=-\frac{\partial^{2} \psi}{\partial x^2}+V(x)\psi,
\label{eq4}
\end{equation}
with $\beta$ denoting the propagation constant. Obviously, the light
propagation is equivalent to a quantum particle in an external
potential.

By applying the Bloch-Floquet theorem, the solutions for
Eq.~(\ref{eq4}) of an infinitely periodic $V(x)$ are Bloch waves,
$\psi_{n,k}(x)=\exp[ikx]u_{n,k}(x)$, where $k$ is the wave number,
$n$ is the band index and $u_{n,k}(x)$ has the same periodicity of
$V(x)$. The Bloch waves of real wave numbers are amplitude-bounded
oscillatory solutions. Otherwise, the Bloch waves of complex wave
numbers show unbounded exponential behavior \cite{Kohn,Heineb}. The
energy spectrum for Eq.~(\ref{eq4}) consists of bands in which there
exist only amplitude-bounded oscillatory solutions and gaps in which
there exist unbounded oscillatory solutions.

Under conditions of $\theta=\arctan(\Delta/(N+1))$, $n_1'=2\eta^2$ and $n_2'=2\eta\sqrt{(N+1)^2+\Delta^2}$ with integers $N\geq0$, the potential $V(x)$ could be denoted by $V(\eta,\Delta,N,x)$ and it supports a set of in-gap solutions (see more details in the Appendix),
\begin{eqnarray}
&&\psi_N^{m_1}(\eta,\Delta,x)= \exp\left[i\left(\frac{N}{2}+i\frac{\Delta}{2}\right)x-2\eta\cos x\right] \nonumber \\
&& \times\sum_{n=0}^N\left [\frac{a_n(\beta_N^{m_1}) }{2}\left(
\exp[-i n x]+
\exp[-i(N-n)x]\right)\right.\nonumber\\
&&\left. +i\frac{b_n(\beta_N^{m_1})}{2}\left(  \exp[-i n
x]-\exp[-i(N-n)x]\right) \right]. \label{solutiona}
\end{eqnarray}
Comparing to the Bloch-wave form, the solution
$\psi_N^{m_1}(\eta,\Delta,x)$ is the Bloch-wave solution  with a
complex wave number $k=N/2+i\Delta/2$.
 Here, $a_n$ and $b_n$ can be derived from two
recursive series: $2\eta(n+2)a_{n+2} +[(n+1)(n+1-N)
+(N^2+\Delta^2)/4 -2\eta^2 -\beta_N^{m_1}]a_{n+1} +2\eta(N-n)a_n
+[(n+1)\Delta-N\Delta/2]b_{n+1}=0$ and $2\eta(n+2)b_{n+2}
+[(n+1)(n+1-N) +(N^2+\Delta^2)/4 -2\eta^2 -\beta_N^{m_1}]b_{n+1}
+2\eta(N-n)b_n +[N\Delta/2-(n+1)\Delta]a_{n+1}=0$ with initial
conditions of $a_0=1$, $b_0=0$,
$a_1=(\beta_N^{m_1}+2\eta^2-(N^2+\Delta^2)/4)/2\eta$ and
$b_1=-N\Delta/4\eta$. The propagation constant $\beta_N^{m_1}$
corresponds to the $m_1$-th real zeros of $d_{N+1}(\eta, \Delta,
\beta)=a_{N+1}+ib_{N+1}=0$ in ascending order. Since $d_{N+1}(\eta,
\Delta, \beta)$ is a polynomial of degree $N+1$ in the propagation
constant $\beta$, $\beta_N^{m_1}$ have at most $N+1$ solutions.
Mathematically, one can construct a linearly independent solution
for $\psi_N^{m_1}$ in form of
$\widetilde{\psi}_N^{m_1}=\psi_N^{m_1}\int_{-\infty}^{x}(\psi_N^{m_1})^{-2}dx$.
Although $\psi_N^{m_1}$ and $\widetilde{\psi}_N^{m_1}$ have the same
propagation constant $\beta_N^{m_1}$, their divergence properties
are opposite: $\psi_N^{m_1} \rightarrow 0$ when
$\widetilde{\psi}_N^{m_1} \rightarrow \infty$ and vice versa.

\begin{figure}[htb]
%\begin{center}
\includegraphics[width=\columnwidth]{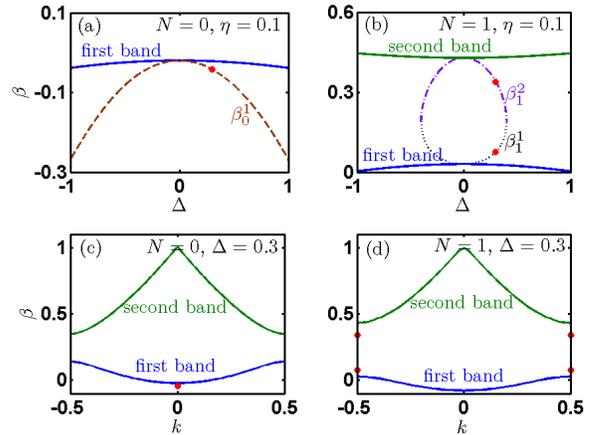}
\caption{(Color online). Band-gap structures for $V(\eta,\Delta, N,
x)$. (a) The first band at $k=0$ and the in-gap propagation constant
$\beta_0^1$ for $N=0$ and $\eta=0.1$. (b) The first two bands at
$k=\pm 0.5$ and the in-gap propagation constants $\beta_1^{1,2}$ for
$N=1$ and $\eta=0.1$. (c) The first two bands and $\beta_0^1$ for
$N=0$ and $\Delta=0.3$. (d) The first two bands and $\beta_1^{1,2}$
for $N=1$ and $\Delta=0.3$. In which, the red dot in (a) corresponds
to the one in (c), and the two red dots in (b) correspond to the
four red dots in (d).} \label{fig1}
%\end{center}
\end{figure}

In principle, for a given superlattice $V(\eta,\Delta,N,x)$, one may
determine the coefficients for the finite-superposition solutions
from the two recursive series. For $N \leq 3$, one can easily obtain
the exact forms for $a_n$ and $b_n$. However, if $N>3$, it is very
difficult to give the exact forms for $a_n$ and $b_n$, and one has
to find their values by using numerical methods. Below, we consider
the two simplest cases: $N=0$ and $N=1$. For the case of $N=0$, the
sole finite-superposition solution is expressed as
\begin{equation}
\psi_0^1(\eta,\Delta, x) = \exp\left[-\frac{\Delta}{2}x -2\eta\cos x\right],
\end{equation}
with $\beta_0^1(\eta,\Delta) = -(\Delta^2+8\eta^2)/4$. For the case
of $N=1$, there are two finite-superposition solutions. The first
finite-superposition solution is in form of
\begin{eqnarray}
&& \psi_1^1(\eta,\Delta,x) =\exp\left[-\frac{\Delta}{2}x -2\eta\cos x\right]\nonumber\\
&& \times
\left[\frac{4\eta-\sqrt{16\eta^2-\Delta^2}}{4\eta}\cos\left(\frac{x}{2}\right)
-\frac{\Delta}{4\eta}\sin\left(\frac{x}{2}\right) \right],
\end{eqnarray}
with
$\beta_1^{1}(\eta,\Delta)=(1-\Delta^2-8\eta^2-2\sqrt{16\eta^2-\Delta^2})/4$.
The other finite-superposition solution reads as
\begin{eqnarray}
&& \psi_1^2(\eta,\Delta, x)=\exp \left[-\frac{\Delta}{2}x -2\eta\cos x\right]\nonumber\\
&& \times
\left[\frac{4\eta+\sqrt{16\eta^2-\Delta^2}}{4\eta}\cos\left(\frac{x}{2}\right)
-\frac{\Delta}{4\eta}\sin\left(\frac{x}{2}\right) \right],
\end{eqnarray}
with
$\beta_1^{2}(\eta,\Delta)=(1-\Delta^2-8\eta^2+2\sqrt{16\eta^2-\Delta^2})/4$.
If $16\eta^2-\Delta^2=0$, the two in-gap waves
$\psi_1^1(\eta,\Delta, x)$ and $\psi_1^2(\eta,\Delta, x)$ are
identical.

In comparison with the band-gap structure, if $\Delta \neq 0$, we
find that $\beta_0^1$ falls into the semi-infinite gap below the
lowest band and $\beta_1^{1,2}$ lies in the first band-gap, see
Fig.~\ref{fig1}. This means that these finite-superposition
solutions are a kind of in-gap states. If $\Delta=0$, the
finite-superposition solutions become stable Bloch-wave solutions,
since the wave numbers become real. Interestingly, $\beta_1^{1}$ and
$\beta_1^{2}$ form a closed circle connecting the first two bands at
$\Delta=0$, see Fig.~\ref{fig1}(b). Moreover, the in-gap state
$\psi_0^1$ appears at $k=0$, while the in-gap states $\psi_1^{1,2}$
appear at $k=\pm 0.5$, see Fig.~\ref{fig1}(c) and (d). Since the
in-gap states grow without bound, they are unphysical states for the
infinite periodic system. However, as we will show below, the in-gap
states can be used to construct  a special class of  exact surface
states in several typical models.

\section{Surface states in single-interface systems}

\begin{figure}[htb]
%\begin{center}
\includegraphics[width=\columnwidth]{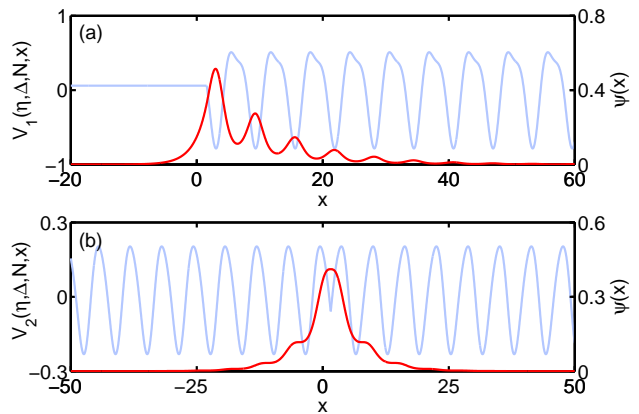}
\caption{(Color online). Surface states in single-interface systems.
(a) $V_1(\eta,\Delta,N,x)$ of $x_0=\pi/2$, $\eta=0.3$, $\Delta=0.2$,
$V_0=0.06$ and $N=0$. (b) $V_2(\eta,\Delta,N,x)$ of $x_0=\pi/2$,
$\eta=0.1$, $\Delta=4\eta\sin x_0$ and $N=0$.}\label{fig2}
%\end{center}
\end{figure}

One of the most famous single-interface systems is a semi-infinite periodic system
of a truncated $V(\eta,\Delta,N,x)$ connecting a constant refractive
index $V_0$~\cite{Heinea, Forstmann, Levinea, Steslicka}. The
potential for such a system reads as,
 \begin{equation}
V_1(\eta,\Delta,N,x) = \left\{ {\begin{array}{lr}
   {V_0,\;\;\;\;\;\;\;\;\;\;\;\;\;\;\;\;\;\; x \leq x_0}\;\; (\textrm{region I}),\\
   {V(\eta,\Delta,N,x), \; x > x_0} \;\;(\textrm{region II}).
\end{array}} \right.\nonumber
\end{equation}
For an allowed surface state, in the region II, it should be in form
of $\psi_N^{m_1}(\eta,\Delta,x)$ for $\Delta>0$ (or
$\widetilde{\psi}_N^{m_1}$ for $\Delta<0$),
$\psi_\text{II}(x)=C_2\psi_N^{m_1}$ (or
$\psi_\text{II}(x)=C_2\widetilde{\psi}_N^{m_1}$). Below we will only
consider the case of $\Delta>0$. In the region I,
$\psi_\text{I}(x)=C_1\exp[\sqrt{V_0-\beta_N^{m_1}}x]$ if
$V_0>\beta_N^{m_1}$. The coefficients $C_i$ are determined by the
normalization condition. By applying the continuity condition at the
interface $x=x_0$, we find that
\begin{equation}
\sqrt{V_0-\beta_N^{m_1}(\eta,\Delta)} = W_N^{m_1}(\eta,\Delta,x_0),
\end{equation}
with $W_N^{m_1}(\eta,\Delta,x) = \dot{\psi}_N^{m_1}(\eta,\Delta,x)
/\psi_N^{m_1}(\eta,\Delta,x)$. Here, the dot denotes the derivative
with respect to $x$. Thus the surface state exists if
$W_N^{m_1}(\eta,\Delta,N,x_0)>0$ and
$V_0=\beta_N^{m_1}+(W_N^{m_1})^2$. For example, in the simplest case
of $N=0$, the interface parameters $x_0$ and $V_0$ satisfy the
conditions,  $\sin(x_0)>\Delta/4\eta$ and
$V_0=-(\Delta^2+8\eta^2)/4+(\Delta/2-2\eta\sin(x_0))^2$.  In
Fig.~\ref{fig2}(a), we show the surface wave in this semi-infinite
periodic system with $\eta=0.3$, $\Delta=0.2$ and $N=0$, which
corresponds to $n_1=2.2\times 10^{-4}$, $n_2=7.6\times 10^{-4}$) and
$\theta=\arctan(0.2)$ in an experimental system of $\lambda=980$ nm,
$n_s=1.518$ and $\Lambda=8$ $\mu$m~\cite{Della, Shandarova}. In this
situation, we take $x_0=\pi/2$ and thus have
$V_0=-(\Delta^2+8\eta^2)/4+(\Delta/2-2\eta\sin(x_0))^2=0.06$.

Another typical single-interface system is of two truncated periodic
potentials connecting at the interface. We consider a system of
$V(-\eta,-\Delta,N,x)$ and $V(\eta,\Delta,N,x)$ with $\Delta>0$
connecting at the interface $x=x_0$~\cite{Goto}. Its potential is
expressed as
\begin{equation}
V_2(\eta,\Delta, N, x) = \left\{ {\begin{array}{lc}
   {V(-\eta,-\Delta,N,x),\; x \leq x_0}\;(\textrm{region I}),\\
   {V(\eta,\Delta,N,x), \;\;\;\;\;\;x > x_0}\;(\textrm{region II}).
\end{array}} \right. \nonumber
\end{equation}
Thus, we have $\psi_\text{I}(x)=C_1\psi_N^{m_1}(-\eta,-\Delta,x)$
for $x\leq x_0$ and
$\psi_\text{II}(x)=C_2\psi_N^{m_1}(\eta,\Delta,x)$ for $x>x_0$. The
continuity conditions at $x=x_0$ give
\begin{equation}
W_N^{m_1}(-\eta,-\Delta,N,x_0)=W_N^{m_1}(\eta,\Delta,N,x_0).
\end{equation}
 For
the case of $N=0$, it is easy to find $4\eta\sin x_0=\Delta$ from
the continuity condition at the interface. Therefore, the surface
wave exists only when $\left|\Delta/4\eta\right|\le 1$. In
Fig.~\ref{fig2}(b), we show the surface state for $N=0$ and
$x_0=\pi/2$. We also find that the position of the interface $x_0$
affects strongly the shape of the surface waves. For an example, the
surface wave for the case of $x_0=\pi/2$ is symmetric about
$x=x_0=\pi/2$. While the surface wave for the case of $x_0=\pi/6$ is
asymmetric about $x=x_0=\pi/6$.

\section{Surface states in double-interface systems}

\begin{figure}[htb]
%\begin{center}
\includegraphics[width=\columnwidth]{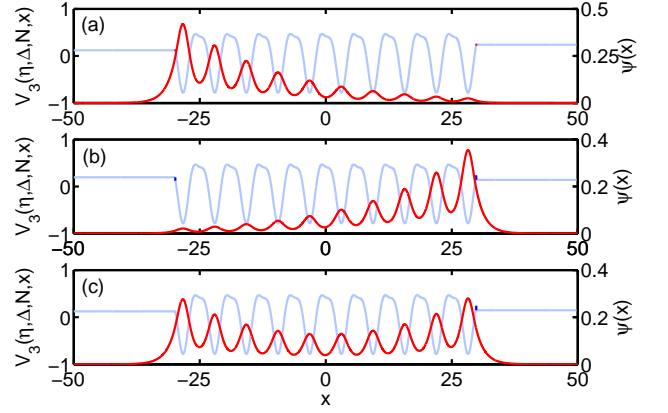}
\caption{(Color online). Surface states in a finite periodic system
described by $V_3(\eta,\Delta,N,x)$ with $N=0$, $\eta=0.3$,
$\Delta=0.1$, and $x_0=-x_1=-19\pi/2$. (a) $V_0$=0.12, $V_1=0.24$,
and $R=0$. (b) $V_0=0.196151$ and $V_1=0.142676$, and $R=\infty$.
(c) $V_0=0.123884$,  $V_1=0.147421$, and $R=0.0696$.}\label{fig3}
%\end{center}
\end{figure}

One typical double-interface system is a finite periodic system
$V(\eta,\Delta,N,x)$ sandwiched by two constant refractive indices
$V_0$ and $V_1$~\cite{Shockley}, which obeys the potential
\begin{equation}
V_3(\eta,\Delta,N, x) = \left\{ {\begin{array}{lc}
   {V_0, \;\;\;\;\;\;\;\;\;\;\;\;\;\;\;\;\;x\leq x_0}\;(\textrm{region I}),\\
   {V(\eta,\Delta,N,x), x_0<x < x_1}\;(\textrm{region II}),\\
   {V_1, \;\;\;\;\;\;\;\;\; \;\;\;\;\;\;\;\;\;x\geq x_1}\;(\textrm{region III}).
\end{array}} \right. \nonumber
\end{equation}
In the region II, one may use the finite-superposition solution
$\psi_N^{m_1}(\eta,\Delta, x)$ and its linearly independent solution
$\widetilde{\psi}_N^{m_1}$ to construct the surface state, that is,
$\psi_\text{II}(x)=C_2\psi_N^{m_1}(\eta,\Delta, x) +
\widetilde{C}_2\widetilde{\psi}_N^{m_1}$. In other two regions, the
physical state must be non-divergent and normalizable. Therefore, if
$V_{0,1}>\beta_N^{m_1}$, we have
$\psi_\text{I}(x)=C_1\exp[\sqrt{V_0-\beta_N^{m_1}}x]$ for $x \leq
x_0$ and $\psi_\text{III}(x)=C_3\exp[-\sqrt{V_1-\beta_N^{m_1}}x]$
for $x\geq x_1$. Similarly, the continuity conditions at the two
interfaces $x=x_0$ and $x=x_1$ request
\begin{eqnarray}
\sqrt{V_1-\beta_N^{m_1}}=-\frac{W_N^{m_1}(\eta,\Delta,x_1)+R\cdot
K_N^{m_1}(x_1)}{1+R\cdot F_N^{m_1}(x_1)},
\end{eqnarray}
with
\begin{eqnarray}
R=\widetilde{C}_2/C_2=\frac{\sqrt{V_0-\beta_N^{m_1}}-W_N^{m_1}(\eta,\Delta,x_0)}{K_N^{m_1}(x_0)-\sqrt{V_0-\beta_N^{m_1}}F_N^{m_1}(x_0)},
\nonumber
\end{eqnarray}
$K_N^{m_1}(x)=\dot{\widetilde{\psi}}_N^{m_1}(x)/\psi_N^{m_1}(x)$ and
$F_N^{m_1}(x)=\widetilde{\psi}_N^{m_1}(x)/\psi_N^{m_1}(x)$. In
principle, the ratio between $C_2$ and $\widetilde{C}_2$ can be
arbitrary. However, to satisfy the continuity conditions at the two
interfaces, the interface parameters $(V_0, V_1, x_0, x_1)$ and the
lattice parameters $(\eta, \Delta, N)$ should obey some certain
conditions. In Fig.~\ref{fig3}, we show two surface states for
$\eta=0.3$, $\Delta=0.1$, $N=0$ and $x_0=-x_1=-19\pi/2$. If $R=0$,
the continuity conditions request $V_0=0.12$ and $V_1=0.24$ and the
surface state is localized around the interface $x=x_0$, see
Fig.~\ref{fig3}(a). If $R=\infty (C_2=0)$, the continuity conditions
request $V_0=0.196151$ and $V_1=0.142676$ and the surface state is
localized around the interface $x=x_1$, see Fig.~\ref{fig3}(b). If
$R=0.0696$, the continuity conditions request $V_0=0.123884$ and
$V_1=0.147421$ and the surface state is almost equally localized
around $x=x_0$ and $x=x_1$, see Fig.~\ref{fig3}(c).

\begin{figure}[htb]
\includegraphics[width=\columnwidth]{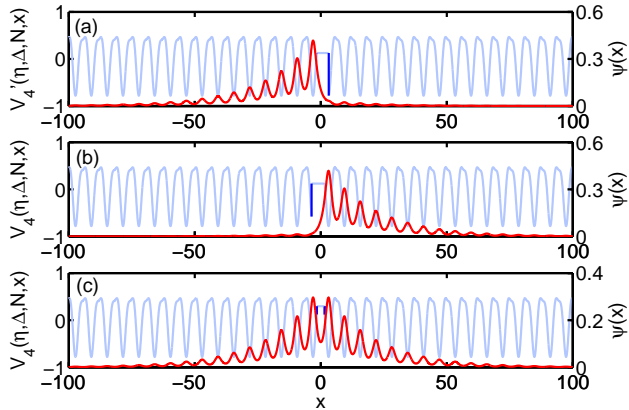}
\caption{(Color online). Surface states in a double-interface system
described by $V_4(\eta,\Delta,N,x)$ with $\eta=0.3$, $\Delta=0.1$
and $N=0$. (a) $x_0=-\pi/2$,  $x_1=\pi$,  $V_0=0.13$, and
$R=0.0255412$. (b) $x_0=-7\pi/6$,  $x_1=\pi/2$,  $V_0=0.120845$, and
$R=254.28$. (c) $x_0=-x_1=-\pi/2$,  $V_0=0.30$, and
$R=1.0$.}\label{fig4}
\end{figure}

Another typical double-interface system is a constant refractive
index $V_0$ sandwiched by two truncated periodic systems
$V(\eta,-\Delta,N, x)$ and $V(\eta,\Delta,N, x)$~\cite{Vinogradov,
Lenz,Ihm}. The corresponding refractive index profile is in form of
\begin{equation}
V_4(\eta,\Delta,N,x) = \left\{ {\begin{array}{lc}
   {V(\eta,-\Delta,N, x), \;\;x \leq x_0}\;(\textrm{region I}),\\
   {V_0,\;\;\;\;\;\;\;\;\; \;\;\;x_0<x < x_1}\;(\textrm{region II}),\\
   {V(\eta,\Delta,N, x),\;\;\;x \geq x_1}\;(\textrm{region III}).
\end{array}} \right. \nonumber
\end{equation}
Thus, we have $\psi_\text{I}(x)=C_1\psi_N^{m_1}(\eta,-\Delta,x)$ in the region I, $\psi_\text{II}(x)=C_2^{-}\exp[-\sqrt{V_0-\beta_N^{m_1}}x] +C_2^{+}\exp[+\sqrt{V_0-\beta_N^{m_1}}x]$ in the region II, and $\psi_\text{III}(x)=C_3\psi_N^{m_1}(\eta,\Delta,x)$ in the region III. The continuity conditions request
\begin{eqnarray}
\frac{W_N^{m_1}(\eta,\Delta,x_1)}{\sqrt{V_0-\beta_N^{m_1}}} = \frac{R\cdot\exp\left[2\sqrt{V_0-\beta_N^{m_1}}x_1\right]-1} {R\cdot\exp\left[2\sqrt{V_0-\beta_N^{m_1}}x_1\right]+1},
\label{conditione}
\end{eqnarray}
with
\begin{eqnarray}
R=\frac{C_2^{+}}{C_2^{-}}&=&\frac{\sqrt{V_0-\beta_N^{m_1}} +W_N^{m_1} (\eta,-\Delta,x_0)}{\sqrt{V_0-\beta_N^{m_1}}-W_N^{m_1} (\eta,-\Delta,x_0)}\nonumber\\ &&\times\exp{[-2\sqrt{V_0-\beta_N^{m_1}}x_0]}.\nonumber
\end{eqnarray}
In Fig.~\ref{fig4}, we show two  surface states for $\eta=0.3$,
$\Delta=0.1$ and $N=0$. The surface waves strongly depend on the two
interface positions. For $x_0=-\pi/2$ and $x_1=\pi$ [or
$x_0=-7\pi/6$ and $x_1=\pi/2$], the surface state has an asymmetric
distribution, see Fig.~\ref{fig4}(a) [or (b)]. While for
$x_0=-\pi/2$ and $x_1=\pi/2$, it becomes symmetric, see
Fig.~\ref{fig4}(c).

\section{Conclusion}

In conclusion, by using the superpositions of finite numbers of
unstable Bloch states for the corresponding infinite periodic
systems, we have given an efficient  approach for constructing a
special class of exact surface states in photonic superlattices.
These exact surface states have the same propagation constants for
the finite-superposition states in the energy gaps and so that they
are a kind of stable in-gap states. This  method has been used to
find parts of the exact surface states in several typical systems
involving a single or two interfaces. By matching two solutions (a
finite-superposition solution and a free-space solution or two
finite-superposition solutions) at two sides of the interfaces, the
existence conditions for surface states are obtained analytically
from the continuity conditions. The existence and the shapes of the
exact surface states not only depend on the interface parameters,
but also rely on the lattice parameters. Our results give an
analytical demonstration of the existence of the surface states and
should shine light on understanding and controlling the surface
waves.

\section*{Acknowledgments}

The authors acknowledge Yuri S. Kivshar for his valuable comments.
This work is supported by the NBRPC under Grant No. 2012CB821300 (2012CB821305), the NNSFC under Grants No. 10905019 and No. 11075223, the PCSIRT under Grant No. IRT0964, the NCETPC under Grant No. NCET-10-0850, the Construct Program of the National Key Discipline and the Fundamental Research Funds for Central Universities of China.

\appendix*
\section{Derivation of  the Bloch-wave solutions $\psi_N^{m_1}(\eta,\Delta,x)$}
Below, we give more details about how to derive the in-gap solutions
\begin{eqnarray}
&&\psi_N^{m_1}(\eta,\Delta,x)= \exp\left[i\left(\frac{N}{2}+i\frac{\Delta}{2}\right)x-2\eta\cos x\right] \nonumber \\
&& \times\sum_{n=0}^N\left [\frac{a_n(\beta_N^{m_1}) }{2}\left(
\exp[-i n x]+
\exp[-i(N-n)x]\right)\right.\nonumber\\
&&\left. +i\frac{b_n(\beta_N^{m_1})}{2}\left(  \exp[-i n
x]-\exp[-i(N-n)x]\right) \right]. \label{solution}
\end{eqnarray}
for the Schr\"{o}dinger equation
\begin{equation}
-\frac{d^2}{dx^2}\psi(x)+V(x)\psi(x)=\beta\psi(x),\label{schrodinger}
\end{equation}
with
\begin{eqnarray}
V(x)&=&-n_1' \cos (2x)+n_2' \cos (x+\theta)\nonumber\\
&=&-n_1' \cos (2x)+n_2'\cos \theta \cos x-n_2'\sin \theta \sin x.\nonumber
\end{eqnarray}

In general, the in-gap solutions stay at the edges of Brillouin
zones and the imaginary parts of their wave numbers are continuous
in a finite region~\cite{Kohn,Heineb}. To find the explicit
expression for a specific set of in-gap solutions, we apply the
following transformation
\begin{eqnarray}
\xi&=&\exp[-ix],\\
\psi(x)&=&\exp[-\sqrt{2n_1'}\cos x]\xi^{\lambda}\phi(\xi),
\end{eqnarray}
with
\begin{eqnarray}
\lambda=\frac{\sqrt{2n_1'}-n_2'\cos \theta-in_2'\sin
\theta}{2\sqrt{2n_1'}}\label{wave}.\nonumber
\end{eqnarray}
Apparently, the parameter $-\lambda$ denotes a complex wave number.
From the Schr\"{o}dinger equation (A.2), we have
\begin{eqnarray}
&&\xi^2\frac{d^2\phi}{d\xi^2}+\left[\sqrt{2n_1'}-\sqrt{2n_1'}\xi^2 +(2\lambda+1)\xi \right] \frac{d\phi}{d\xi}\nonumber\\
&&+\left[\lambda^2-\beta-n_1'-\left(\sqrt{2n_1'}-n_2'\cos\theta\right)\xi\right]\phi=0.\nonumber\\ \label{heund}
\end{eqnarray}

By writing the solution for $\phi(\xi)$ as a standard power-series expansion,
\begin{equation}
\phi(\xi)=\sum_{n=0}^\infty d_n \xi^n
\end{equation}
one can easily find that the coefficients $d_n$ are determined by a three-term recurrence relation,
\begin{equation}
c_0(n)d_n+c_1(n)d_{n+1}+c_2(n)d_{n+2}=0,\label{recurrence}
\end{equation}
with the initial condition $d_0=1$ and $d_{-1}=0$, where
\begin{eqnarray}
c_0(n)&=&n_2'\cos \theta-\sqrt{2n_1'}(n+1),\nonumber\\
c_1(n)&=&(n+1)(n+2\lambda+1)+\lambda^2-n_1'-\beta,\nonumber\\
c_2(n)&=&\sqrt{2n_1'}(n+2).\nonumber
\end{eqnarray}

Usually, the solution (A.6) is an infinite series. However, similar to the procedure of obtaining the Hermite polynomials for a harmonic oscillator, one can impose the truncation condition $d_j=0$ with $j\geq N+1$ and then the solution (A.6) becomes a finite series. If $d_{N+1}=0$ and $d_{N+2}=0$, we can get all following $d_{j}=0$ with $j>N+2$.

To obtain $d_{N+1}=0$, from the three-term recurrence relation (\ref{recurrence}) with $n=N-1$, we have
\begin{equation}
c_0(N-1)d_{N-1}+c_1(N-1)d_{N}=0. \label{conditiona}
\end{equation}
For a given periodic lattice, due to $d_{N}$ is a polynomial of degree $N$ in $\beta$, this equation requests that the propagation constant $\beta$ must be a solution for a polynomial of degree $N+1$ in $\beta$.

To obtain $d_{N+2}=0$, from the three-term recurrence relation (\ref{recurrence}) with $n=N$, the coefficient $c_0(N)$ should satisfy
\begin{equation}
c_0(N)=n_2'\cos \theta-\sqrt{2n_1'}(N+1)=0. \label{conditionb}
\end{equation}
Clearly, this equation requires a special relation between the lattice parameters $n_1'$, $n_1'$ and $\theta$.

Therefore, under the conditions (\ref{conditiona}) and (\ref{conditionb}), the series solution $\phi(\xi)$ becomes a polynomial,
\begin{equation}
\phi_N(\xi)=\sum_{n=0}^N d_n(\beta_N^{m_1}) \xi^n.
\end{equation}
After some mathematical calculation, we get the following in-gap solution
\begin{eqnarray}
\psi(x)&=&\exp {\left[i\left(\frac{N}{2}+i\frac{n_2'\sin \theta}{2\sqrt{2n_1'}}\right)x -\sqrt{2n_1'}\cos x\right]}\nonumber\\
&&\times \sum_{n=0}^N d_n(\beta_N^{m_1})\exp{\left[-inx\right]}.
\end{eqnarray}
Here, $\beta_N^{m_1}$ denotes $m_1$-th real zero of $d_{N+1}=0$ in
ascending order. It is clear that $\psi(x)$ are Bloch-wave solutions
with the complex wave numbers $k=-\lambda=N/2+i\frac{n_2'\sin
\theta}{2\sqrt{2n_1'}}$. If $\sqrt{2n_1'}=2\eta$, $n_2'\cos
\theta=2\eta(N+1)$ and $n_2'\sin\theta=2\eta \Delta$, we have
$n_1'=2\eta^2$, $n_2'=2\eta\sqrt{(N+1)^2+\Delta^2}$ and
$\theta=\arctan(\Delta/(N+1))$. The complex wave number is given as
\begin{equation}
k=\frac{N}{2}+i\frac{\Delta}{2}.
\end{equation}

Due to $V(x)$ is a real function, we can take the real part of $\psi(x)$ as a solution of Eq.~(\ref{schrodinger})
\begin{eqnarray}
&&\psi_N^{m_1}(\eta,\Delta,x)= \exp\left[i\left(\frac{N}{2}+i\frac{\Delta}{2}\right)x-2\eta\cos x\right] \nonumber \\
&& \times\sum_{n=0}^N\left [\frac{a_n(\beta_N^{m_1}) }{2}\left(
\exp[-i n x]+
\exp[-i(N-n)x]\right)\right.\nonumber\\
&&\left. +i\frac{b_n(\beta_N^{m_1})}{2}\left(  \exp[-i n
x]-\exp[-i(N-n)x]\right) \right].\nonumber\\
\end{eqnarray}
where $d_n(\beta_N^{m_1})=a_n(\beta_N^{m_1})+ib_n(\beta_N^{m_1})$. This completes the derivation of Bloch-wave solutions~(\ref{solution}).

In the following, we show how to give the in-gap waves for the cases
of $N=0$ and $N=1$. For the case of $N=0$, from
Eq.~(\ref{conditiona}) with $N=0$, we have
\begin{eqnarray}
c_1(-1)d_0&=&\lambda^2-n_1'-\beta \nonumber\\
          &=&-\frac{\Delta^2}{4}-2\eta^2-\beta=0.
\end{eqnarray}
This equation has only one real root $\beta_0^1=-(\Delta^2+8\eta^2)/4$. From Eq.~(\ref{solution}) with $N=0$, the corresponding in-gap state is given as
\begin{equation}
\psi_0^1(\eta,\Delta, x) = \exp\left[-\frac{\Delta}{2}x -2\eta\cos
x\right].
\end{equation}
For the case of $N=1$, from Eq.~(\ref{conditiona}) with $N=1$, we have
\begin{equation}
c_0(0)d_{0}+c_1(0)d_{1}=0.
\end{equation}
Given $c_0(0)=2\eta$,$c_1(0)=-(\Delta+i)^2/4-2\eta^2-\beta$ and $d_1=(\beta+(\Delta-i)^2/4+2\eta^2)/2\eta$, we have
\begin{equation}
(\beta+(\Delta+i)^2/4+2\eta^2)(\beta+(\Delta-i)^2/4+2\eta^2)-4\eta^2=0.
\end{equation}
This equation have two real roots,
$\beta_1^{1}(\eta,\Delta)=(1-\Delta^2-8\eta^2-2\sqrt{16\eta^2-\Delta^2})/4$
and
$\beta_1^{2}(\eta,\Delta)=(1-\Delta^2-8\eta^2+2\sqrt{16\eta^2-\Delta^2})/4$.
For $\beta_1^{1}(\eta,\Delta)$, we have $a_0=1$, $b_0=0$,$a_1=-\sqrt{16\eta^2-\Delta^2}/4\eta$, and $b_1=-\Delta/4\eta$, therefore the corresponding in-gap state reads as
\begin{eqnarray}
&& \psi_1^1(\eta,\Delta,x) =\exp\left[-\frac{\Delta}{2}x -2\eta\cos x\right]\nonumber\\
&& \times
\left[\frac{4\eta-\sqrt{16\eta^2-\Delta^2}}{4\eta}\cos\left(\frac{x}{2}\right)
-\frac{\Delta}{4\eta}\sin\left(\frac{x}{2}\right) \right].\nonumber\\
\end{eqnarray}
Similarly, for $\beta_1^{2}(\eta,\Delta)$, we have $a_0=1$, $b_0=0$,$a_1=\sqrt{16\eta^2-\Delta^2}/4\eta$, and $b_1=-\Delta/4\eta$. The corresponding in-gap state reads as
\begin{eqnarray}
&& \psi_1^2(\eta,\Delta,x) =\exp\left[-\frac{\Delta}{2}x -2\eta\cos x\right]\nonumber\\
&& \times
\left[\frac{4\eta+\sqrt{16\eta^2-\Delta^2}}{4\eta}\cos\left(\frac{x}{2}\right)
-\frac{\Delta}{4\eta}\sin\left(\frac{x}{2}\right) \right].\nonumber\\
\end{eqnarray}

\end{document}